\documentclass[aps,pre,twocolumn,amsmath,amssymb,showpacs,showkeys,floatfix]{revtex4-2}
\usepackage{epsfig}
\usepackage{graphicx}
\usepackage{setspace}
\usepackage{bm}
\usepackage{float}
\usepackage{amsmath}
\usepackage{braket}
\usepackage{xcolor}
\usepackage[colorlinks,bookmarks=false,citecolor=blue,linkcolor=red,urlcolor=blue]{hyperref}

\begin{document} 

\title{Asymmetric dynamical localization and precision measurement of BEC micromotion}
\author{S. Sagar Maurya, J. Bharathi Kannan, Kushal Patel, Pranab Dutta, Korak Biswas}
\affiliation{Department of Physics, Indian Institute of Science Education and Research, Pune 411008, India}
\author{M. S. Santhanam}
\email{santh@iiserpune.ac.in}
\affiliation{Department of Physics, Indian Institute of Science Education and Research, Pune 411008, India}
\email {Electronic mail: santh@iiserpune.ac.in}
\author{Umakant D. Rapol}
\email{umakant.rapol@iiserpune.ac.in}
\affiliation{Department of Physics, Indian Institute of Science Education and Research, Pune 411008, India}

\date{\today}

\begin{abstract}
We employ a Bose Einstein Condensate (BEC) based atom-optic kicked rotor to generate an asymmetrically localized momentum distribution that depends upon initial velocity of the BEC. Asymmetric features are shown to arise from the early-time dynamics induced by the broken parity symmetry and, asymptotically freeze as the dynamical localization stabilizes. The asymmetry in the momentum distribution critically depends upon the initial launch velocity and is sensitive to very small initial velocities (`micromotion') of the BEC.  In this work, we also perform a precise measurement of the `micromotion'. By utilizing the technique of measuring the early-time asymmetry of momentum distribution, we report measurement of micromotion down to 230 $\pm$17  $\mu$m/s.
\end{abstract}
\maketitle
\section{Introduction}
In recent years, there has been much interest in studying quantum kicked rotor (QKR) with Bose-Einstein condensates (BEC) -- both with and without tunable interactions \cite{PhysRevLett.110.190401,toh2023evidence,SeeToh2022,Cao2022}. Further, the ease of imprinting the lattice phase on the BEC also leads to different types of effects such as on-resonance quantum ratchets \cite{PhysRevLett.100.024103, PhysRevLett.99.043002} and quantum boomerang effect \cite{PhysRevX.12.011035,PhysRevA.103.063316,PhysRevB.105.L180202}. BEC based QKR has also been utilized in coupled quantum kicked rotors \cite{PhysRevLett.110.190401} where two incommensurate optical lattices drive a quantum-to-classical transition by breaking dynamical localization (DL). Further,  a discrete-time quantum walk has also been observed in a periodically kicked BEC-based system \cite{PhysRevLett.121.070402}, and a variant with spin-1/2 degree of freedom has been proposed to realize and detect topological phase transitions \cite{PhysRevA.106.043318,PhysRevResearch.5.043167}. In this work, we utilize quantum kicked rotor experimentally realized with BEC in optical lattices for precision measurement of initial velocities towards applications in atom interferometry.

Precision measurements with atom interferometers have opened up tremendous applications in quantum sensing \cite{Bongs2019, PhysRevLett.111.143001, PhysRevLett.117.138501}. Atom interferometers have been successfully utilized in gravimetry \cite{Peters2001, PhysRevA.88.043610,PhysRevLett.117.138501}, rotation sensing \cite{PhysRevLett.114.063002, PhysRevLett.116.183003}, magnetometers \cite{PhysRevA.82.061602}, and the determination of photon recoil \cite{PhysRevLett.106.080801}.  In particular, in precision rotation sensing and gravimetry,
nullifying the systematic shifts and measurement errors (arising from Coriolis effect \cite{PhysRevLett.108.090402})  necessitates precise knowledge of the initial velocity \cite{PhysRevLett.114.063002,hardman2016bec,Peters2001}. This initial velocity may arise from externally induced launch velocities or micromotion -- small movement of the atomic cloud -- during the turnoff of the trapping potential. In a rotation sensor, the phase shift caused due to this initial velocity adds to the Sagnac phase shift \cite{PhysRevLett.108.090402}. 
The Sagnac phase shift is in the range of tens of milliradians (mrad) for a velocity of 100 $\mu$m/s over an interferogram time of hundreds of milliseconds (ms). The velocity of micromotion can lie in the range of 100-1000 $\mu$m/s, which is much smaller compared to the velocity imparted due to two-photon recoil momentum ($\approx$ 1 cm/s). The measurement of such a small velocity is challenging because it requires huge time-of-flight in standard absorption imaging or very precise Bragg or Raman spectroscopy to measure the Doppler shift \cite{hardman2016bec,PhysRevLett.82.4569,PhysRevA.65.013403}.
 
In this work, we experimentally perform a precision measurement of `micromotion' of the BEC. Firstly, we experimentally demonstrate asymmetric localization in momentum space by launching the BEC with varying initial velocity or by inducing the lattice motion. The accumulated phase difference imprinted on the launched wave packet or the lattice motion creates an asymmteric momentum distribution. Additionally, we illustrate that the asymmetric nature of the momentum distribution can be  controlled by adjusting the launch velocity rather than altering the direction of the launch velocity. In contrast to previous studies \cite{Jones2004} with cold atoms, where such asymmetry arises from pulse shape effects and observed after long time, here we demonstrate that such asymmetry arises much earlier in time and freezes due to dynamical localization. By utilizing the asymmetric behavior of the momentum distribution in the early-time dynamics, a method is presented to measure micromotion of the BEC \cite{PhysRevA.101.013619, Sun2023, Mangaonkar_2020} in QKR system. The QKR had already been utilized for measurement of gravity through survival resonances \cite{PhysRevA.98.063614,PhysRevE.85.036205} and this work adds another such example of precision measurement using BEC-based QKR.

\section{QKR in moving frame of reference}

The QKR is a fundamental model of quantum chaos \cite{IZRAILEV1990299} extensively explored for its chaotic dynamics and for demonstration of dynamical localization \cite{SanPauKan22, MooRobBha94}. The classical kicked rotor, when strongly kicked by an external field, can display chaotic dynamics accompanied by diffusive growth of mean energy with time. Over extended time scales, quantum interference effects inhibit classical diffusive dynamics, a phenomenon in momentum space analogous to Anderson localization \cite{FisGrePra82}. Since the first realization of the QKR using cold atoms \cite{MooRobBha95}, it has spurred numerous experimental investigations to explore a variety of scenarios that manipulate localization \cite{JonGooMea07, KenGonPat08, PhysRevE.106.034207}. Physically, the standard kicked rotor describes a particle subjected to periodic kicks imparted by the stationary optical lattice created by counter-propagating laser beams. However, in this work, we consider a kicked rotor system in which the optical lattice moves at a constant velocity in the laboratory frame. The Hamiltonian of QKR in a moving lattice is given by \cite{PhysRevA.97.061601}
\begin{equation}
H=\frac{\widehat{p}^2}{2} + K\cos(2k \widehat{x}-2\pi\alpha t) ~ \sum_{n=1}^{N} ~ \delta(t-nT),
\end{equation}
where, $\widehat{p}$ and $\widehat{x}$ represent the momentum and position operators respectively. They obey canonical commutation relation $[\widehat{x}, \widehat{p}] = i \hbar_{\rm eff} $, where the effective Planck constant $\hbar_{\rm eff}$ can be tuned in the experiment. Further, $K$ is the stochastic parameter, $T$ is the time period between consecutive kicks, $k$ is the wave vector, and $\alpha$ is the frequency difference between two lattice beams that make up the optical lattice. The lattice velocity arising from the frequency difference between the counter-propagating beams is  $v = \frac{\lambda\alpha}{2}$, where $\lambda$ is the wavelength of the optical lattice. Throughout this work, parameters will be chosen so that the classical analogue of QKR displays chaos, with $K \geq 5$. This parameter choice ensures that the localization effects we observe are of quantum origin. 

\begin{figure}
	\includegraphics*[width=0.95\linewidth]{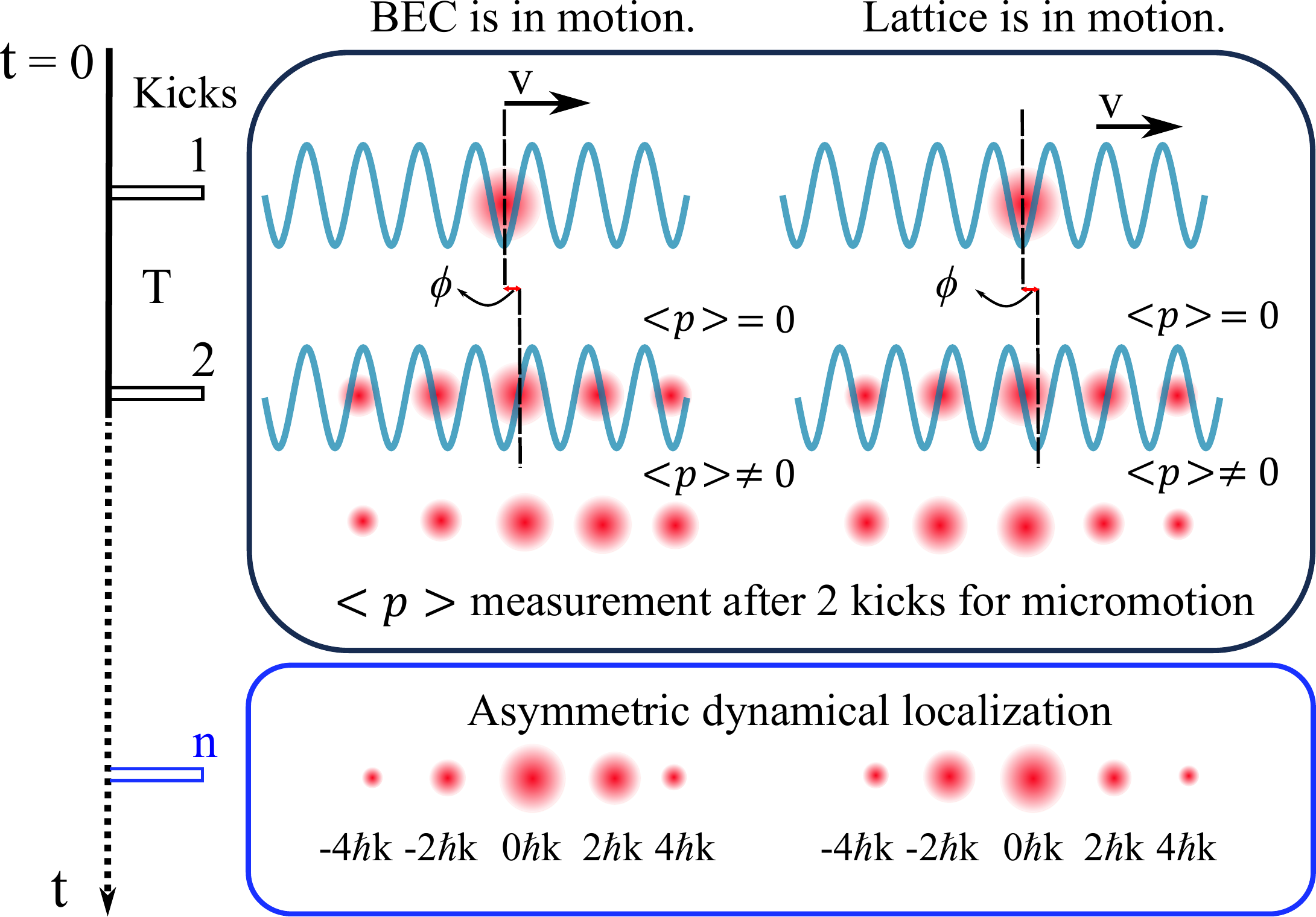}
	\caption{This experimental schematic shows the moving BEC (left side) with velocity $v$ and periodically kicked 1D optical lattice. This relative motion between BEC and lattice provides asymmetry in momentum distribution which we quantify by measuring $\langle p \rangle$. Same asymmetry with opposite nature arises when lattice is in motion, shown in right side. Micromotion is measured by scanning lattice velocity (see Fig. \ref{fig:5} for details). Further, dynamical localization sets in for larger number of kicks.}
	\label{fig:1}
\end{figure}

As the quantum kicked rotor is time-periodic, the quantum dynamics can be conveniently analyzed through the period-1 Floquet operator
\begin{equation}
U = \exp{\left(-i \frac{p^{2}T}{2}\right)} \exp{ \left(-i K \cos(2kx-2\pi\alpha t)\right) }.
\end{equation}
This evolves an initial state  $\psi(x,t=0)$ over one kick period $T$, {\it i.e.}, $\psi(x,T) = U \psi(x,0)$ with the initial state chosen as a coherent state in position space given by
\begin{equation}
    \psi(x,t=0) = \frac{1}{\sqrt{2\pi}\sigma_{w}} \exp\left(-\frac{x^{2}}{2\sigma_{w}^2} \right) \exp(-ip_{0}x),
\end{equation}
and is consistent with the initial distribution of BEC.
Here, $\sigma_w$ characterizes the width of the wavefunction in position space, while $p_{0}$ is the initial velocity, typically arising from launched velocity or micromotion. Generally, in our experiments, when BEC is launched with initial momentum $p_{0}$, the lattice velocity is stationary and vice versa. In general, either moving the lattice or the BEC in same direction are expected to induce similar effects in early time dynamics with exactly opposite asymmetric momentum distribution \cite{Lepers2011, PhysRevLett.100.024103}. The experimental protocol is illustrated in Fig. \ref{fig:1}.

For numerical simulations, we use the standard split-operator method to evolve an initial state of the kicked rotor. This method consists of two primary components: the kick operator, which is diagonal in position space, and the free evolution operator, which is diagonal in momentum space. To simulate scenarios where the lattice is in motion in the lab frame, we imprint the phase just before each kick. Similarly, to model the situation in which BEC moves in the laboratory frame, a wavefunction is initialized incorporating the motion right from the outset.

\subsection*{Asymmetric dynamical localization and micromotion}
If the lattice is moved with constant velocity, then the kick potential (in the lab frame) $V(x) = K\cos(2k x - 2\pi\alpha t)$ induces a path difference at each kick. Consequently, the phase difference between successive kicks is  $\phi = 2\pi\alpha T$ and this breaks the parity symmetry. The total phase difference accumulated after $n$ kicks is $\phi_{n} = 2\pi\alpha T \times (n-1)$, assuming that the phase is initialized to zero for the first kick. The accumulated phase difference $\phi_n$ over short timescales induces an inhomogeneity in position space due to broken parity symmetry, leading to an asymmetric momentum distribution \cite{PhysRevLett.100.024103}. This asymmetry can be quantified through $\langle p(t=2T)\rangle$ immediately following $n=2$ kicks. 

In the experiments, the lattice velocity $v$ is controlled by tuning $\alpha$ and subsequently measuring $\langle p(t=2T)\rangle$. Based on physical consideration and since phases are unique only upto $2\pi$, we posit that the average momentum after two kicks to have a form
\begin{eqnarray}
	\langle p (t=2T) \rangle  = c ~\sin[4\pi(vT/\lambda)],
 \label{eq:a}
\end{eqnarray}
where $c$ is a constant. Exactly the same expression holds good for moving BEC with opposite sign in average $\langle p \rangle$. Further, Eq. \ref{eq:a}  implies that if $vT/\lambda = n/4$ (where $n$ is an integer),  $\langle p \rangle = 0$ implying an absence of asymmetry for specific choice of initial velocity and kick period. In the long time limit of $n \gg 1$, the initial asymmetry accumulated in the short-time limit eventually freezes due to the emergence of dynamical localization. Hence, the early-time dynamics dictates the long-term behavior and the onset of asymmetrical dynamical localization in the system.

To gauge micromotion accurately, the optical lattice is precisely moved with frequency difference of the order of $100$Hz, aligning it with the scale of micromotion. Upon achieving a velocity for the lattice that corresponds precisely to the micromotion, we observe  that $\langle p(t=2T)\rangle$ equals to zero. This alignment establishes a direct correspondence between the velocity of the lattice and the velocity of micromotion.
 
\section{Experimental setup for QKR}
The QKR setup we have used for this work is similar to the one described in Ref. \cite{PhysRevE.106.034207}. However, instead of cold atoms, we utilize a BEC of $^{87}$Rb every 8 seconds, through forced evaporative cooling. The atoms are initially prepared in the $\ket{F=1, m_{F}=-1}$ state, with a BEC temperature of $80$ nK and a population of approximately $40,000$ atoms. This BEC  serves as the initial wave function for our experimental investigations. The optical standing wave is produced using two independent laser beams, which are generated by a single laser passing through two separate acousto-optic modulators. The frequency difference between lattice beams and their switching can also be controlled. Switching time and laser power provide us control over scaled Planck constant $\hbar_{\rm eff}$ and stochastic parameter $K$.

This work comprises two main components: probing asymmetric dynamical localization and measuring micromotion. To investigate localization phenomena, we implement the Bragg diffraction technique to launch the Bose-Einstein condensate with varying recoil momentum, as outlined in Ref. \cite{PhysRevLett.107.130403}. By adjusting the frequency difference between the lattice beams and the on-time of the lattice beam, we achieve a good transfer of atoms to different momentum states. In our apparatus \cite{Dutta2023}, a frequency difference of $15$ kHz results in $2$ photon recoil momenta, while a frequency difference of $30$ kHz provides $4$ photon recoil momenta to atoms, both have a fixed on-time of approximately $\sim 66.6$ $\mu$s. For a given frequency difference, optical lattice moves with half of the speed of diffracted wave-packets. After the creation of the initial wave function of BEC with different velocities, we apply the usual kicked rotor pulse sequence to study the dynamical localization. In our QKR experiments, we maintain the stochastic parameter at $K=5$ and effective Planck constant at $\hbar_{\rm eff}=4.6$ to ensure that the corresponding classical dynamics remains in the chaotic regime \cite{SeeToh2022}.

For the motion of the optical lattice, we generate a frequency difference ranging from $0$ to $75$ kHz between the lattice beams. This frequency range corresponds to a velocity range:
\begin{equation}
	v = \frac{\alpha}{15kHz}.\frac{\hbar k}{M_{\text{Rb}}},
\end{equation}
where, $\hbar$k is single photon recoil momentum. 
We ensure that the kick strength remains constant across higher frequency regimes. The advantage of employing a moving lattice, rather than a moving BEC, lies in the flexibility to assign precise arbitrary velocities to the optical lattice from the laboratory frame. It also does not create any residual atoms in zeroth momentum state like in Bragg diffraction. Leveraging this control over micromotion, we scan the lattice velocity in a frequency range of $-3$ to $3$ kHz with increments of  $100$ Hz, enabling precise measurement of micromotion BEC by balancing the relative motion.

\section{Asymmetric dynamical localization in moving frame of reference}
\label{sec:Asymmetric dynamical localization in moving frame of reference}
In this section, we will consider two scenarios -- (a) BEC launched with an initial momentum in a stationary lattice (called case-I), (b) BEC launched with zero initial momentum in a moving lattice (called case-II).

\subsubsection{Case I: BEC moving in lab frame}
In our experiment, BEC is launched with various recoil velocity, $v_n = 2n \hbar k/M_{Rb}$ ( $n \in \mathbb{Z}$ ), using Bragg diffraction, where $\hbar k/M_{Rb}$ is single photon recoil velocity. This is achieved by applying a pulse of length approximately $66.6$ $\mu$s to transfer all the population to the required momentum state, and appropriately adjusting the lattice power. Subsequently, free evolution period of approximately $66.6$ $\mu$s is allowed, corresponding to the Talbot time of the system \cite{PhysRevLett.83.5407}. This ensures recreation of the initial wavefunction without any unintended phase accumulation \cite{PhysRevA.94.043620}. Once the wavefunction with different velocities is created, it is subjected to periodic kicks to observe dynamical localization. The period of these kicks is set at $T=24.3$ $\mu$s, corresponding to a scaled Planck constant of $\hbar_{\rm eff}=4.6$. The stochastic parameter is $K=5$ corresponding to a classical phase space that is almost fully chaotic \cite{PhysRevE.106.034207,SeeToh2022}. These parameters remain consistent throughout the paper unless stated otherwise.

\begin{figure}
	\centering
	\includegraphics*[width=1.05\linewidth]{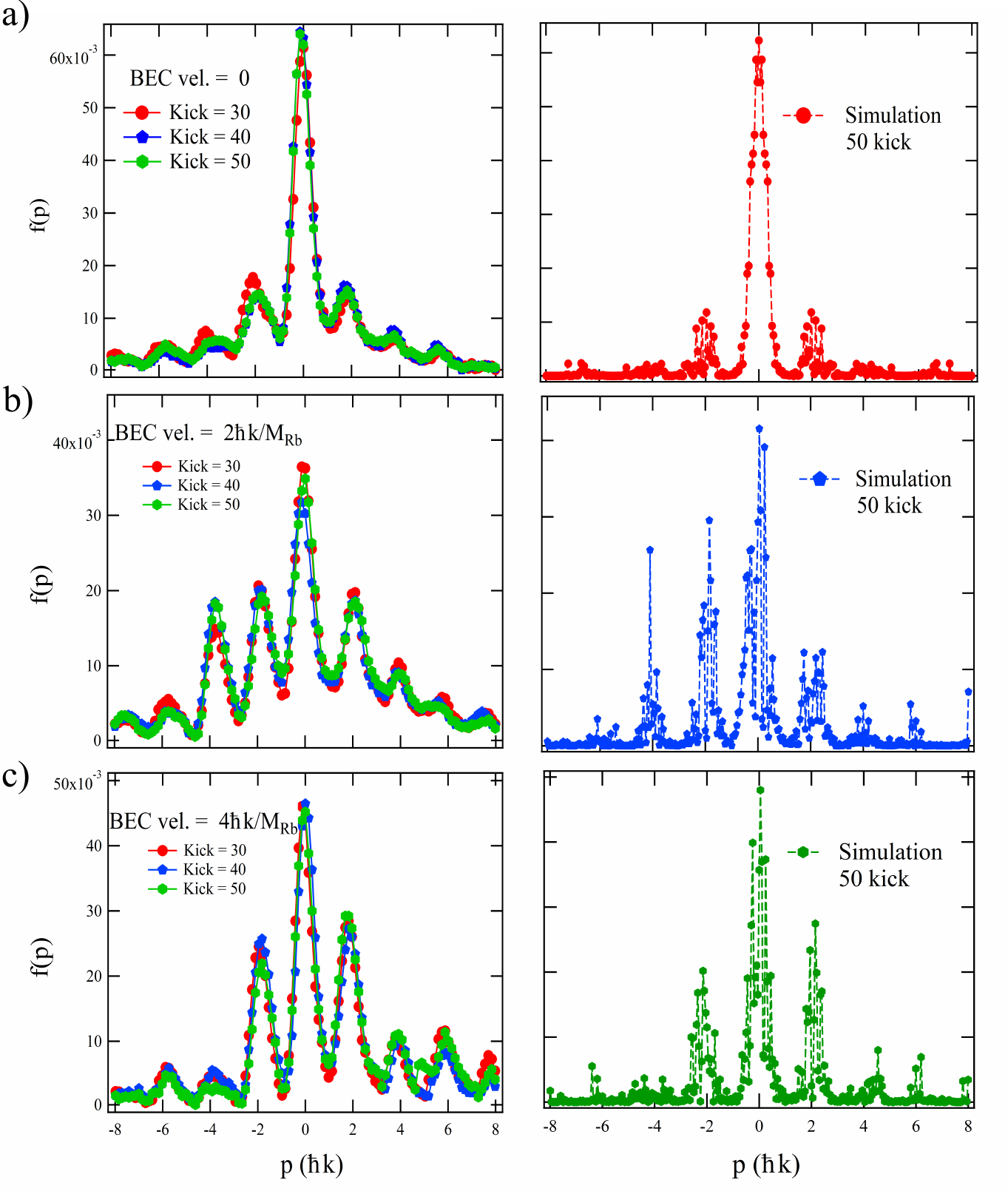}
	\caption{Wavefunction profile for different launch velocity shown at $30$, $40$ and $50$ kicks. In all the cases, dynamical localization has set in for initial velocity corresponds to momenta (a) $0 \hbar k/M_{Rb}$, (b) $2 \hbar k/M_{Rb}$, and (c) $4\hbar k/M_{Rb}$. In all cases, in left side, symbols are obtained experimental data. In right side, symbols are obtained from kicked rotor simulations with 50 kicks.}
	\label{fig:2}
\end{figure} 

Bose-Einstein Condensate is launched with initial velocity of $v_0 =0, 2\hbar k/M_{Rb}$ and $4\hbar k/M_{Rb}$. The momentum distributions observed after $30, 40$, and $50$ kicks are depicted in Fig. \ref{fig:2}(a-c) respectively in the left panel. In Fig. \ref{fig:2}(a), standard symmetric dynamical localization pattern is observed for initial velocity $v_0=0$. Figure \ref{fig:2}(b), illustrates an asymmetric dynamically localized momentum distribution when BEC is launched with $v_0=2\hbar k/M_{Rb}$ (peaked at $2\hbar k$, shifted to zero for better comparison). Similarly, in Fig. \ref{fig:2}(c), dynamical localization is slightly asymmetric for a launch velocity of $v_0=4\hbar k/M_{Rb}$( peaked at $4\hbar k$, shifted to zero for better comparison). Asymptotically, as $n \gg 1$, the system remembers the initial velocity $v_0$ since the maxima of the steady-state distribution occurs at $p=|p_0|$. In Fig. \ref{fig:2}(b), $\langle p \rangle$ with reference to initial given velocity $v_0$, is moving in the direction of the launched velocity (left direction), while in Fig. \ref{fig:2}(c), small $\langle p \rangle$ is moving in the opposite direction of the launched velocity. As for the velocity of $v_0=4\hbar k/M_{Rb}$, $\langle p \rangle$ is small from the beginning, difficult to observe the asymmetry in localized state.

The right panel in Fig. \ref{fig:2} shows the corresponding results obtained from QKR simulations. The simulation results confirm the emergence of dynamical localization and evidently it is asymmetric for the case when BEC is launched with $v_0=2\hbar k/M_{Rb}$ and $v_0=4\hbar k/M_{Rb}$. For $v_0=2\hbar k/M_{Rb}$, the more population lies in the direction of the launched velocity and for $v_0=4\hbar k/M_{Rb}$, it lies in opposite direction with little asymmetry, matching with the experimental result. This asymmetry feature in early-time dynamics will be discussed further in upcoming section. The experimental profile in the vicinity of the peak value, in an average sense, is slightly elevated compared to simulation results (in both Figs. \ref{fig:2}-\ref{fig:3}) in due to presence residual thermal atoms.

\subsubsection{Case II: optical lattice moving in lab frame}
\begin{figure}
	\centering
	\includegraphics*[width=1.01\linewidth]{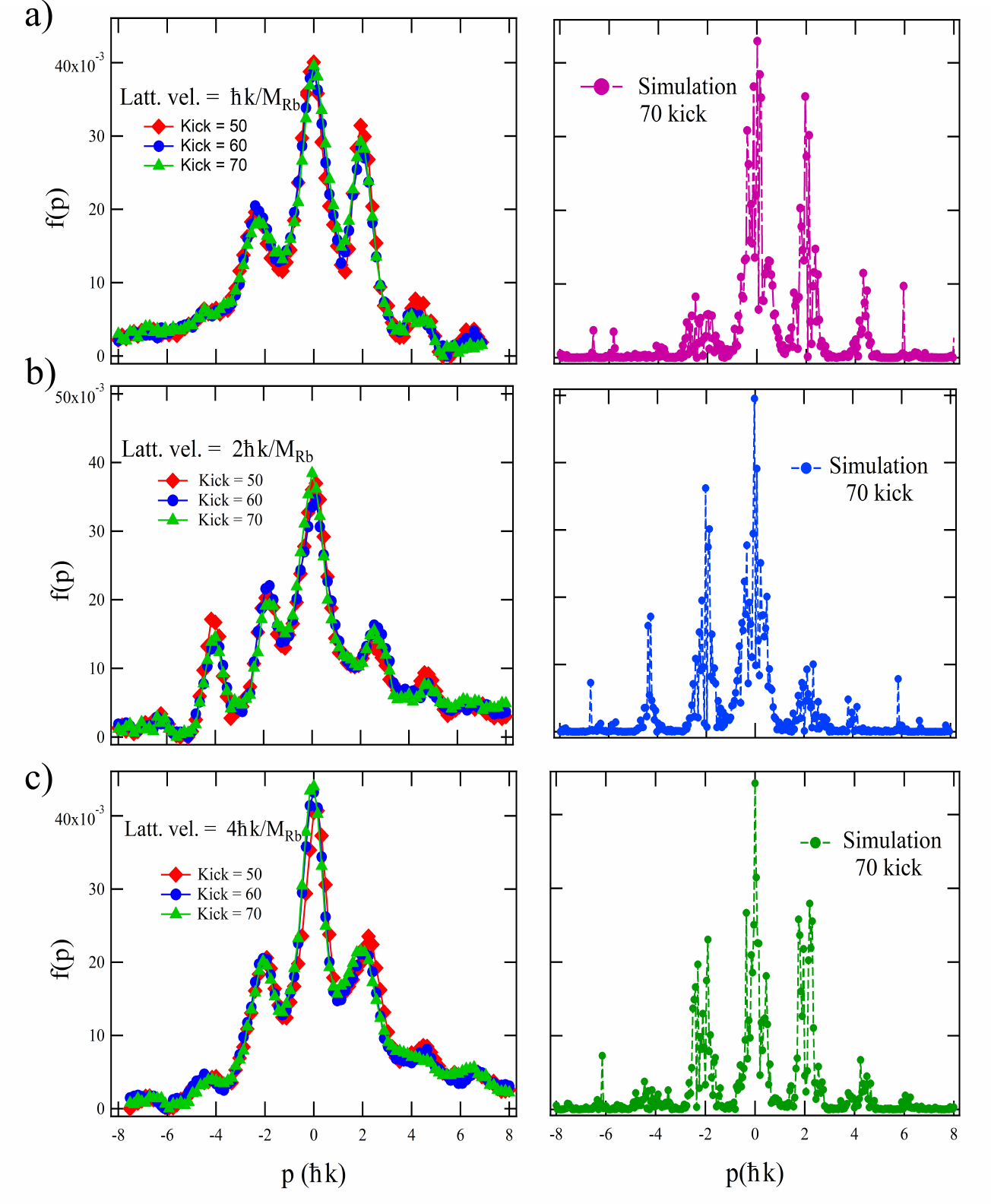}
	\caption{Dynamically localization for three different lattice velocity: (a) $\hbar k/M_{Rb}$, (b) $2 \hbar k/M_{Rb}$ and (c) $4 \hbar k/M_{Rb}$. (left side) experimental data shown as symbols for three different kick numbers, (right side) simulation results shown as symbols at 70th kick. }
	\label{fig:3}
\end{figure}

In this section, we discuss the dynamical localization by moving the lattice in the lab frame by inducing a constant frequency difference $\alpha$ in the range of $0-75$ kHz, which spans almost $5$ recoil velocity. After creating the BEC, $100 \mu s$ time-of-flight is allowed, and subsequently periodic kicks are applied. As seen in Figs. \ref{fig:3}(a-c), dynamical localization is observed for various lattice velocities.
For a direct comparison with case-I, the lattice has been moved in same direction, the observed distribution in this case has been flipped as discussed in Eq. \ref{eq:a} . The optical lattice is moved with velocity $\hbar k/M_{Rb}$ and $2 \hbar k/M_{Rb}$ and $4 \hbar k/M_{Rb}$ in the left direction. The resulting steady-state momentum distribution after $50$, $60$ and $70$ kicks is displayed in Figs. \ref{fig:3}
(a-c). The asymmetry induced by the lattice motion is visible for $\hbar k/M_{Rb}$, $2\hbar k/M_{Rb}$ and $4 \hbar k/M_{Rb}$; for $\hbar k/M_{Rb}$ in Fig. \ref{fig:3}(a) the asymmetric wave packets are moving in the opposite direction as lattice (left direction), for $2 \hbar k/M_{Rb}$ in Fig. \ref{fig:3}(b), the asymmetric wave packets are moving in the same direction as lattice (left direction), and for $4 \hbar k/M_{Rb}$ in Fig. \ref{fig:3}(c), small asymmetric wave packet is moving again in the opposite direction as lattice. 

The right panel in Fig. \ref{fig:3} shows the corresponding localization pattern obtained from QKR simulations. The simulation results confirm the emergence of asymmetric dynamical localization when the optical lattice is moved by velocities of $\hbar k/M_{Rb}$, $2\hbar k/M_{Rb}$ and $4\hbar k/M_{Rb}$.

\section{Measurement of early time dynamics}
\begin{figure}
	\centering
	\includegraphics*[width=1\linewidth]{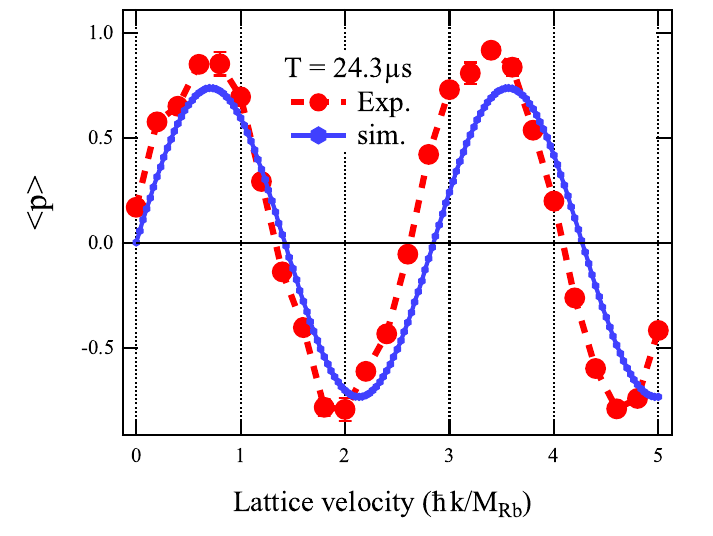}
	\caption{Oscillatory behavior of $\langle p \rangle$ as a function of lattice velocity. In this experiment, $\langle p \rangle$ is measured after two kicks}
	\label{fig:4}
\end{figure}

To understand asymmetry induced by relative motion between the atomic cloud and the optical lattice, early-time dynamics after two kicks is analyzed. The optical lattice is moved by creating a frequency difference between two laser beams, ranging from 0 to $5\hbar k/M_{Rb}$. Subsequently, two kicks are applied separated by time interval $T = 24.3$ $\mu$s for different initial velocity $v$ of lattice and $\langle p \rangle$ is measured after a $10$ ms time-of-flight. Remarkably, pronounced oscillations in $\langle p \rangle$ are observed, consistent with Eq. \ref{eq:a}. This is shown in Fig. \ref{fig:4}, and suggests that $\langle p \rangle$ exhibits a linear relationship in the limit of $v \to 0$. This linearity provides a promising avenue to measure the micromotion discussed in Section~\ref{sec:Measurement of micromotion of BEC}. Fig. \ref{fig:4} also shows simulation results (blue line), which agrees  with the experimental results in the limit $v \to 0$. For the large $v$, where the lattice velocity is high, experiment deviates from the simulation due to finite pulse time.

Another intriguing feature observed in Fig. \ref{fig:4} is the difference in the sign of $\langle p \rangle$ for $\hbar k/M_{Rb}$ and $2 \hbar k/M_{Rb}$. The opposite signs for $\langle p \rangle$ in the early-time dynamics is a consequence of the asymmetric distribution observed after long-time evolution, as demonstrated in Section~\ref{sec:Asymmetric dynamical localization in moving frame of reference}. In general, short time asymptotic carry the signature of the long-term behavior at other velocities as well. In particular, for velocities that are integer multiples of $1.37\hbar k/M_{Rb}$ (observed $1.35\hbar k/M_{Rb}$ from Fig. \ref{fig:4}), asymmetry is absent in the distribution. This corresponds to the condition $vT/\lambda = n/4$, when asymmetry is expected to vanish. Irrespective of the speed of lattice, if the velocity is an integer multiple of $1.37\hbar k/M_{Rb}$, no asymmetry is expected to manifest in the distribution, as a consequence of Eq. \ref{eq:a}.

\section{Measurement of micromotion of BEC}

\label{sec:Measurement of micromotion of BEC}
Measuring the micromotion of the BEC poses a challenge due to its very small magnitude in the direction of the lattice \cite{benaicha2024dual}. This small magnitude of the velocity is difficult to measure accurately using conventional time-of-flight methods, which typically require long time-of-flight duration. One such measurement is performed in Ref. \cite{hardman2016bec} through a $225$ ms time-of-flight. To address this challenge and accurately measure micromotion, early-time measurements are conducted by systematically tuning $\alpha$ (frequency difference) of the lattice. A single-shot measurement can also measure micromotion if the lattice phase is stable. In these measurements, two different time delays, $T = 24.3 \mu$s and $T = 50 \mu$s, between two kicks, are employed.
\begin{figure}
	\centering
	\includegraphics*[width=1\linewidth]{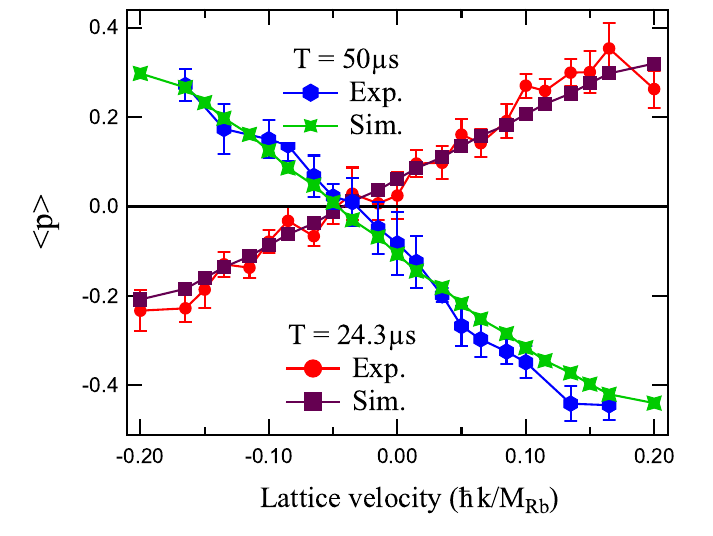}
	\caption{Measured $\langle p \rangle$ as a function of lattice velocity by changing the frequency difference $\alpha$ for two different time delays. Micromotion of BEC is measured by scanning over a range of lattice velocity. See text for details.}
	\label{fig:5}
\end{figure}
The BEC micromotion is measured by tuning the lattice velocity in steps of $0.017\hbar k/M_{Rb}$ corresponding frequency difference of $250$ Hz, as illustrated in Fig. \ref{fig:5}. The phase evolution induced by micromotion is effectively balanced by the lattice motion. When the net average momentum $\langle p \rangle = 0$, the BEC micromotion velocity can be deduced from the corresponding lattice velocity by linear fit. The calculated lattice velocity at points of $\langle p \rangle = 0$ are $(0.039\pm 0.003)\hbar k/M_{Rb}$, corresponding to BEC micromotion velocity of $(230\pm17)$ $\mu$m/s for $T = 24.3$ $\mu$s, and $(0.043\pm 0.002)\hbar k/M_{Rb}$, corresponding to a velocity of $(254\pm10)$ $\mu$m/s for $T = 50$ $\mu$s. The micromotion can also be obtained by measuring asymmetry, keeping the lattice velocity at $v=0$, if the proportionality constant $c$ of Eq. \ref{eq:a} is known. However, the former method is favoured as it offers a direct and precise measurement of the micromotion.

\section{Conclusion}
This work gives insights about the nature of dynamical localization under two scenarios; (i) when a wave packet is launched with initial momenta $p_0 \ne 0$ in the lab frame, and (ii) when the optical lattice is moved in lab frame. In both scenarios, after a short diffusive timescale, the wavepacket is localized in momentum space with an asymmetric distribution profile. This asymmetry emerges during the early time dynamics -- driven by the breaking of parity symmetry due to the motion of the wavepacket or the lattice. This feature is employed for precisely measuring the micromotion of the BEC. For the parameters of the experimental system we employed, velocity measurements yielded $(230\pm17)$ $\mu$m/s for $T = 24.3$ $\mu$s and $(254\pm10)$ $\mu$m/s for $T = 50$ $\mu$s in our system. This micromotion measurement is crucial for precision instruments such as atom interferometers and atomic gyroscopes, to correct for systematic shifts and uncertainties. The micromotion velocity is an order of magnitude smaller than one recoil photon momentum, as well as comparable to mean velocity associated with BEC temperature.  The broken parity symmetry induced by the micromotion is utilized to measure such small velocity. Further, it might not be significantly affected by velocity distribution of the BEC, a common challenge in spectroscopy technique. This work contributes to our understanding of the precision measurement techniques using BEC-based quantum kicked rotor models.

\section{Acknowledgments}
S.S.M. acknowledges research fellowship from Council of Scientific and Industrial Research (CSIR), Government of India. All the authors thank the National Mission on Interdisciplinary Cyber Physical Systems for funding from the DST, Government of India through the I-HUB Quantum Technology Foundation, IISER Pune.

%

\end{document}